# Evolution of magnetism in cerium doped $LaCo_2P_2$ crystals: a magnetic phase diagram

Yong Tian,[1] Yixiu Kong,[1] Kai Liu,[1] Anmin Zhang,[1] Rui He,[2] Qingming Zhang[1,†]

[1] Department of Physics, Beijing Key Laboratory of Opto-electronic Functional Materials & Micro-nano Devices, Renmin University of China, Beijing 100872, People's Republic of China
[2] Department of Physics, University of Northern Iowa, Cedar Falls, Iowa 50614, USA

## Abstract

$ThCr_2Si_2$-type phosphide $ACo_2P_2$ (A=Rare earth elements) has the same structure as iron arsenides, but their magnetic behaviors are quite distinct. In this paper, we for the first time grew a series of $La_{1-x}Ce_xCo_2P_2$ single crystals (x=0.0 to1.0), and made structural and magnetic characterizations. This allows us to carry out a careful investigation on the evolution of magnetism with cerium content and build a magnetic phase diagram. We found that the introduction of cerium induces a rapid decrease of c-axis and a change from ferromagnetic (FM) to antiferromagnetic (AFM) states. By employing first-principles band-structure calculations, we identify the formation of P-P bonding with the shortening of c-axis, which effectively drives an increase of AFM interaction and eventually leads to AFM ordering in the high doping region. The present study may shed light on the interplay between the structural collapsing and electronic/magnetic properties in 122 iron pnictides.

[†] qmzhang@ruc.edu.cn

# Introduction

The discovery of superconductivity in iron-arsenic compounds has generated great interest since 2008[1-3]. The iron-based superconductors of "122" family have become one of the focused research areas due to their simple structures and high crystalline quality [4, 5]. A special issue in the 122 compounds is the distinct magnetism and phase transition of $CaFe_2As_2$ and $EuFe_2As_2$ under pressure[6-10]. At atmospheric pressure, $CaFe_2As_2$ crystal forms collinear AFM order below 170 K. However, under moderate pressure, arsenic atoms in the adjacent layer form As-As bonding[11] which leads to a nonmagnetic collapsed phase[12]. The collapsed phase transition dramatic reduces the c lattice parameter and also exists in CoAs system. On the other hand, CoAs system exhibits different magnetic behaviors. $BaCo_2As_2$[12] and $SrCo_2As_2$[13] are nonmagnetic in the uncollapsed tetragonal (uT) phase -- $CaCo_2As_2$, which has a c-axis of 10.27Å close to the "collapsed tetragonal" (cT) iron arsenic, shows an interlayer AFM ordering[14, 15]. In fact, more than seven hundred $ThCr_2Si_2$-type compounds have been investigated[14-17]. Among them, $AM_2X_2$ (A = bivalent or trivalent ions, M = magnetic ions, X = As or P) has distinct magnetic properties which largely depend on the vertical distance of X ions from the plane of M ions. And they are also affected by many other factors like the sizes of A ions, the valence states of M ions and the substitution of X ions, etc.

It is expected that the substitution of arsenic atoms by smaller phosphorus atoms will bring much richer magnetic properties and the evolution of magnetism will be an interesting issue. As mentioned above, 122 system can be divided into two magnetic categories according to the different lengths of *c* axis. This can also be applied to the CoP family. $EuCo_2P_2$ has an uT phase without magnetic ordering on Co sites[18], while $PrCo_2P_2$, $NdCo_2P_2$, and $CeCo_2P_2$ are in the cT phase[19, 20] with interlayer AFM ordering like $CaCo_2As_2$. More interestingly, intralayer ferromagnetism has been observed in the CoP-based compounds with the substitution of moderate-size rare earth atoms, such as $LaCo_2P_2$ and $CaCo_2P_2$[19, 21], which are intermediate compounds between non-magnetism $EuCo_2P_2$ and antiferromagnetism $CeCo_2P_2$.

In this paper, we report our studies of $La_{1-x}Ce_xCo_2P_2$. We have grown a series of $La_{1-x}Ce_xCo_2P_2$ crystals (x=0.0 to 1.0) and made careful structural characterizations. The magnetization measurements indicate that their magnetism strongly depends on the content of Ce. And the rapid increase of AFM coupling temperatures with Ce content exactly follows the decrease of lattice constants of c-axis, indicating a tight interplay between lattice and magnetism. By combining the results with first-principles calculations, we found that Ce content effectively modulates the distance between P-P atoms, which eventually results in the change of magnetism from interlayer FM to AFM. We further built a magnetic phase diagram, in which the critical Ce content from FM to AFM is located at x ~ 0.65.

**Experiments and methods**

Our starting materials were finely dispersed powders of lanthanum (99.95% General Research Institute for Nonferrous Metals), cerium (99.95% General Research Institute for Nonferrous Metals), cobalt (99.9% Alfa Aesar) and phosphorus (99.9% Alfa Aesar). Tin shots (99.99%) were also obtained from Alfa Aesar. Precursor CoP was synthesized from cobalt and phosphorus power after 48-hour annealing at 1073 K.

The rare-earth cobalt phosphides $La_{1-x}Ce_xCo_2P_2$ ($x \leq 0.4$) were prepared by tin flux synthetic procedure, which can be found elsewhere[22, 23]. For the synthesis of $La_{1-x}Ce_xCo_2P_2$ ($x$=0.65, 0.88, 1), cobalt and phosphorus powder were replaced by the precursor CoP with the same molar ratio. The starting materials were mixed and sealed in silica tubes with air pressure lower than 1 Pa. After that the mixtures were annealed at 1153 K for 7 days. Then they were cooled down to 873 K at a rate of 10 K/min and then quenched in water. The tin shots containing La (Ce) $Co_2P_2$ samples were dissolved in diluted HCL for 48 h. The crystals are the shiny, sheet-like residues with a typical size of $0.3 \times 0.3 \times 0.05$ mm$^3$.

X-ray diffraction was performed at room temperature on a Bruker D8 Advance using CuKα radiation. Elemental analysis of single crystals was conducted using Scanning Electron Microscope-Energy-Dispersive X-ray Spectroscopy (SEM-EDX, FEI NOVA NanoSEM 450). Magnetic susceptibility measurements were carried out using Magnetic Property Measurement System (MPMS III, Quantum Design).

First-principles calculations were performed with the projector augmented wave method[24, 25] as implemented in the Vienna Ab initio Simulation Package (VASP)[26-28]. For the exchange-correlation potential, the generalized gradient approximation (GGA) of Perdew-Burke-Ernzerhof (PBE) type was used. The kinetic energy cutoff of the plane wave basis was set to be 350 eV. A $16\times16\times6$ k-point mesh was employed for the Brillouin zone sampling and the Gaussian smearing with a width of 0.05 eV was adopted around the Fermi surface. Both cell parameters and internal atomic positions were allowed to relax until the forces were smaller than 0.01 eV/Å.

## Results and discussions

Figure 1 shows SEM images of six crystals and the analysis on element ratio. The images demonstrate that our samples are well crystallized and the crystalline growth interface can be identified. The typical size of the sample is hundreds of microns. The nominal and actual $x$ concentrations obtained from EDX are summarized in Table 1. The actual contents of Ce are a little higher than the nominal ones systemically and the former are used in this paper.

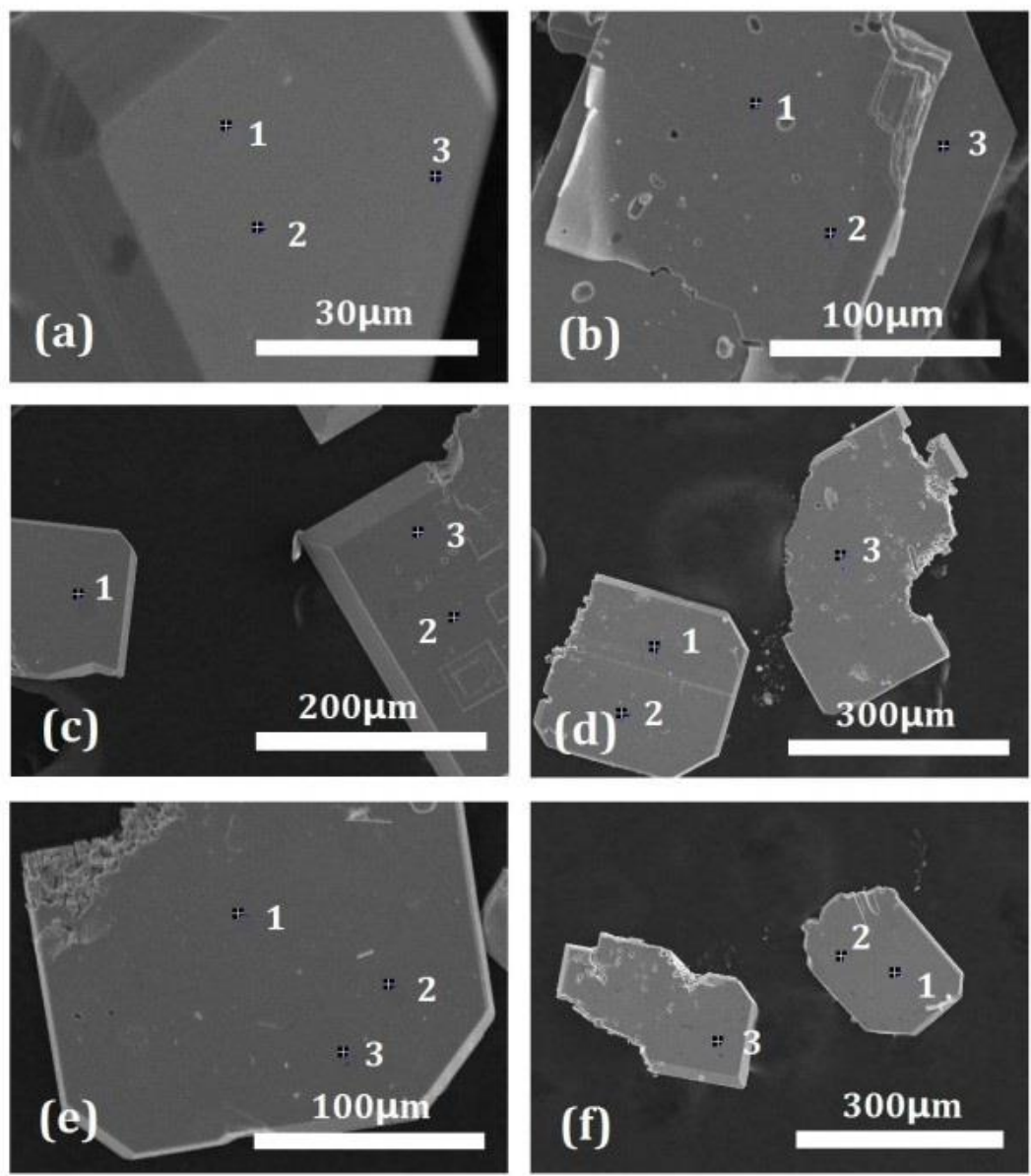

**Figure 1** (a) - (f) SEM images of $La_{1-x}Ce_xCo_2P_2$ crystals with actual content $x$=0, 0.15, 0.35, 0.49, 0.65, 0.88, respectively. EDX was conducted at three points on each sample (marked as 1, 2, 3). The average $x$ values are listed in Table 1.

**Table 1** Nominal and actual contents of $La_{1-x}Ce_xCo_2P_2$ determined by EDX

| Nominal content | $x$=0 | $x$=0.1 | $x$=0.25 | $x$=0.4 | $x$=0.55 | $x$=0.85 | $x$=1 |
| --- | --- | --- | --- | --- | --- | --- | --- |
| Actual content | $x$=0 | $x$=0.15 | $x$=0.35 | $x$=0.49 | $x$=0.65 | $x$=0.88 | $x$=1 |

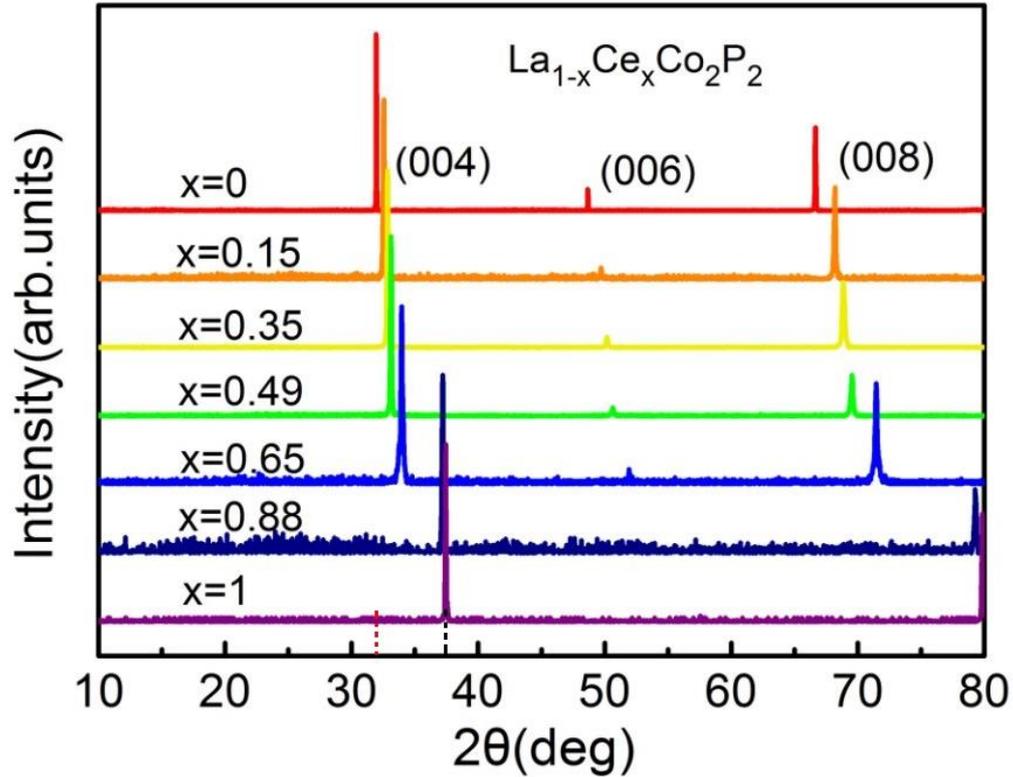

**Figure 2** X-ray diffraction pattern of $La_{1-x}Ce_xCo_2P_2$. Miler indices of three crystal planes are marked next to the peaks. The red and black vertical dashed lines highlight the positions of the 004 peak of $LaCo_2P_2$ and $CeCo_2P_2$, respectively.

Figure 2 shows X-ray diffraction patterns of $La_{1-x}Ce_xCo_2P_2$ single crystals. Only three ($00l$) diffraction peaks for each crystal can be seen, since the measurements were performed on ab plane. All peaks monotonically shift to larger diffraction angles as the Ce content increases. This indicates that La atoms have been successfully substituted by Ce atoms since smaller cerium atoms causes a shrink of the lattice and an increase of diffraction angle $\theta$. A prominent change in Fig. 2 is the abrupt increase of $\theta$ when $x$ changes from 0.65 to 0.85, suggesting a phase transition or a critical point in the doping region. This is well consistent with our magnetic measurements (see below).

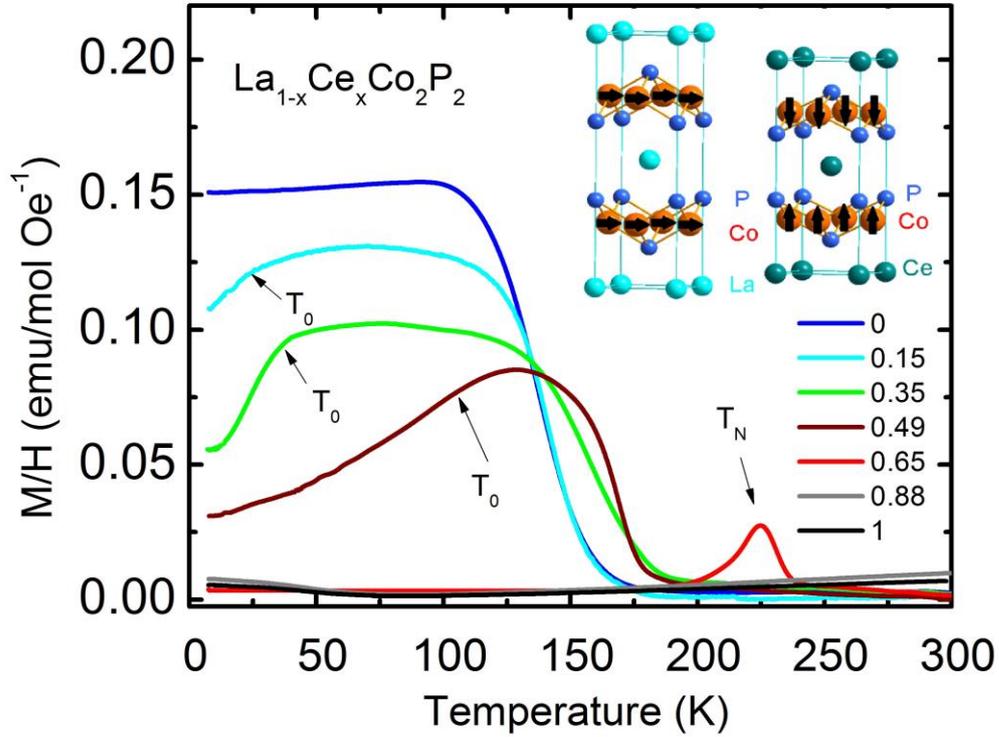

**Figure 3** Magnetic susceptibility of $La_{1-x}Ce_xCo_2P_2$ crystals with different x values. Inset: magnetic structures of $LaCo_2P_2$ and $CeCo_2P_2$.

Figure 3 shows susceptibilities in $La_{1-x}Ce_xCo_2P_2$ crystals. The susceptibility curve for *x*=0 ($LaCo_2P_2$) agrees well with that reported in Ref.[21]. In the samples with $x \leq 0.49$, a change in susceptibility is seen around 150 K, indicative of a ferromagnetic transition at this temperature. Below the ferromagnetic transition temperature, the susceptibility start to decreases as Ce doping into the compounds. Furthermore, the onset temperature of this decrease continues to rise from x=0.15 to 0.49. Spin moments lie in *a-b* plane in $LaCo_2P_2$[21] but form AFM ordering along c axis in $CeCo_2P_2$[29], as seen in the inset of Fig. 3. Doping of Ce results in an enhancement of AFM coupling along the c axis. The decrease of susceptibility is attributed to the new emerge interlay AFM coupling, similar figures are also been observed in La (Pr) $Co_2P_2$[22]. The temperatures of the susceptibility drop are defined as $T_0$, which describes the characteristic temperature of interlayer AFM coupling. As more Ce atoms enter into the lattices, $T_0$ increases and the ferromagnetic component of susceptibilities gradually decreases. When *x* reaches to 0.65, a Curie-Weiss peak emerges, forming AFM order and the ferromagnetism completely disappears. The system transforms from intralayer FM of $LaCo_2P_2$ to interlayer AFM of $CeCo_2P_2$. The $CeCo_2P_2$ (*x*=1) and $La_{0.12}Ce_{0.88}Co_2P_2$ (*x*=0.88) samples show a relative flat susceptibility curve since their AFM transition temperatures (440K for the $CeCo_2P_2$[19]) are out of the temperature range of the present measurements.

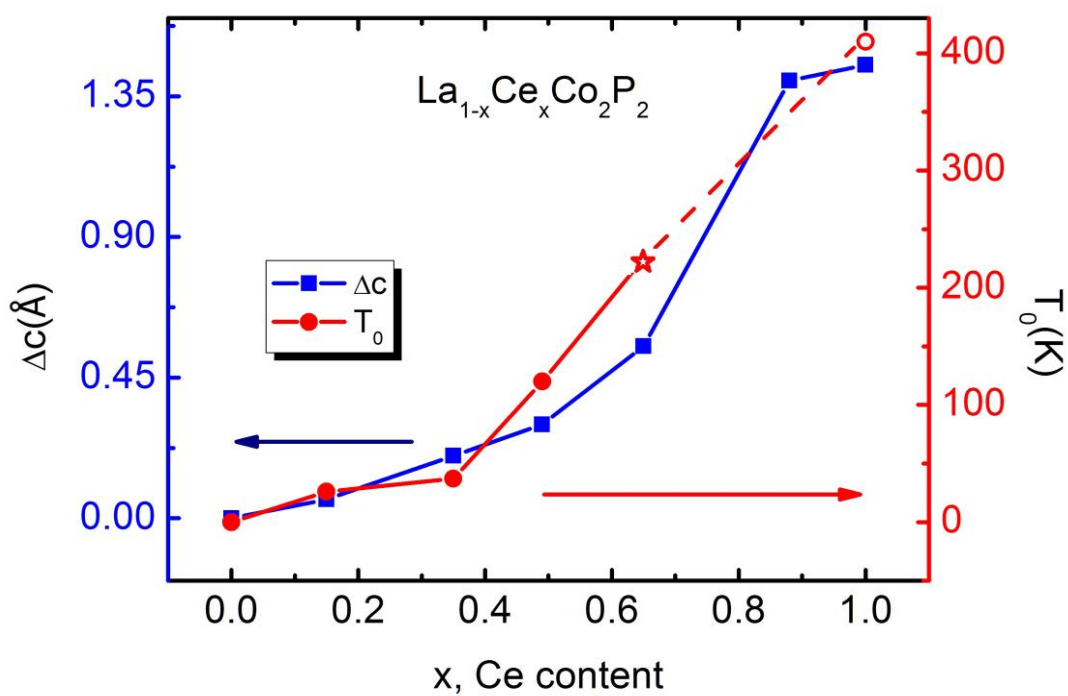

**Figure 4** Ce content dependence of $T_0$ and c-axis lattice constants relative to that of parent compound ($\Delta C = C_{x=0} - C_x$, in which $C_x$ is obtained from XRD shown in Fig. 2, and $C_{x=0}$ =11.04 Å). The red open star and circles are AFM transition temperature $T_N$ and the latter is obtained from ref. 26.

The Ce content dependence of $T_0$, $T_N$ and *c*-axis lattice constants ($\Delta c$) are displayed in Fig. 4. The *c*-axis lattice constant dramatically changes in the range of x = 0.4 - 0.9. Interestingly, the increase of $T_0$, $T_N$ with the increase of Ce content exactly follows the evolution of c-axis lattice constants. The similar dependence suggests a tight connection between magnetism and lattice in the system.

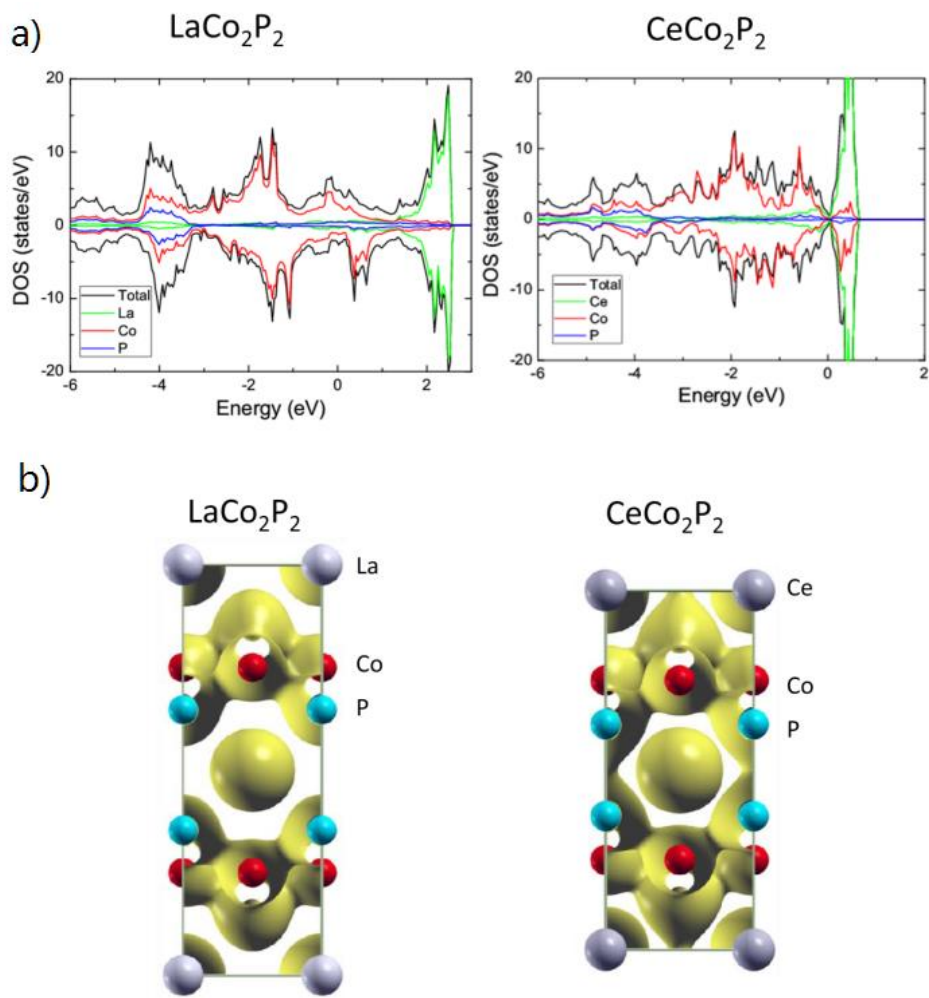

**Figure 5** (a) Electronic density of states and (b) charge densities with the same isosurface value of $LaCo_2P_2$ and $CeCo_2P_2$ in their respective magnetic ground states,

To explore the relation of magnetism and lattice in $La_{1-x}Ce_xCo_2P_2$, we have carried out first-principles calculations for the starting (parent) compound $LaCo_2P_2$ and the end compound $CeCo_2P_2$. The density of states (DOS) for $LaCo_2P_2$ and $CeCo_2P_2$ in their respective magnetic ground states are given in Fig. 5a, the charge densities of $CeCo_2P_2$ (Fig. 5b) show a clear overlap between P atoms from different CoP layers, indicating that the formation of P-P pairs offers an interlayer channel crucial for AFM exchange interaction. This may explain the connection between cT phase and AFM ordering. Actually, in CoP family, the cT compounds tend to be antiferromagnetic while the uT compounds are paramagnetic or ferromagnetic. It is expected that the introduction of smaller Ce atoms enhances AFM coupling and induces the transition into AFM phase. The bonding of nearest neighboring P atoms in adjacent layers is a key to understand the correlation between the collapsed phase and magnetic transition. Hoffmann *et. al.*[30] have shown that the P-P potential has two minima which correspond to the non–interacting and bonding P-P pairs, respectively. This can be distinguished by the distance of P-P atoms. The cT phase compounds ($PrCo_2P_2$[20], $CeCo_2P_2$[29] and $EuCo_2P_2$[31] under pressure) has 2.5 Å P-P distances—close to covalent bonding in phosphorus molecule and tend to form bond. The P-P distance in uT phase compounds ($LaCo_2P_2$[21], $SrCo_2P_2$[29], $EuCo_2P_2$[31] at atmospheric pressure) usually larger than 3.0 Å and exist no interlayer interaction. The shortening of c-axis in $La_{1-x}Ce_xCo_2P_2$ induces a phase transition from uT to cT phases, accompanied by the formation of P-P bonding and interlayer AFM ordering.

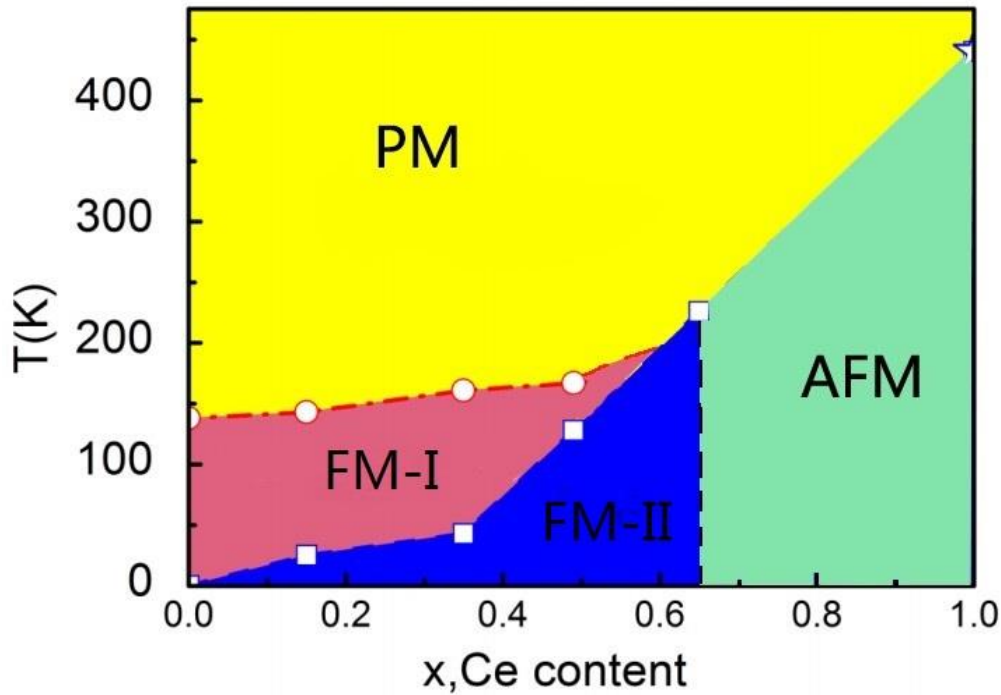

**Figure 6** Magnetic phase diagram of $La_{1-x}Ce_xCo_2P_2$. The star-shaped data point was obtained from ref. 20. PM, FM-I, FM-II, AFM denote the paramagnetic phase, ferromagnetic phase without AFM interaction, ferromagnetic phase with AFM interaction, and antiferromagnetic, respectively.

The results above allow us to build a magnetic phase diagram of $La_{1-x}Ce_xCo_2P_2$ system (see Fig. 6). The parent compounds ($LaCo_2P_2$) shows a ferromagnetic transition at ~125 K (marked as FM-I). As cerium content increases, AFM interaction emerges in the low-temperature FM phase (marked as FM-II), continue to rise with higher characteristic temperature, until the AFM transition appears and the FM order vanishes at x~ 0.65, accompanied by a structural collapse along c axis.

**Conclusion**

In summary, $La_{1-x}Ce_xCo_2P_2$ (x=0, 0.15, 0.35, 0.65, 0.88, 1) single crystals have been grown for the first time and their structural and magnetic properties have been carefully characterized. We found that the doping of cerium induces AFM interaction and a collapsed phase transition, accompanied by an AFM phase transition, occurs at x~0.65. First-principles calculations suggest that the P-P bonding in the cT phase is responsible for the enhancement of interlayer antiferromagnetic interaction. Finally, the results allow us to propose a magnetic phase diagram to summarize the rich magnetic properties in the 122 system.

**Acknowledgements**

This work was supported by the Ministry of Science and Technology of China (Grant No.: 2016YFA0300504) and the NSF of China (No.: 11474357). Q.M.Z., A.M.Z. and K.L were supported by the Fundamental Research Funds for the Central Universities and the Research Funds of Renmin University of China (2112009382, 14XNLF06, 16XNH063).